\documentclass[prl,twocolumn,preprintnumbers,superscriptaddress]{revtex4-1}
\usepackage{amssymb}
\usepackage{amsmath}
\usepackage{amsfonts}
\usepackage{graphicx}
\usepackage{dcolumn}
\usepackage{bm}
\usepackage{verbatim}
\usepackage{color}
\usepackage{wasysym}
\usepackage{xcolor}
\usepackage{ulem}

\begin{document}

\title{Effective potentials for polar molecules under non-orthogonal 
dual microwave fields}
\date{\today}

\author{Fulin Deng}
\affiliation{Institute of Theoretical Physics, Chinese Academy of Sciences, Beijing 100190, China}

\author{Xinyuan Hu}
\affiliation{Institute of Theoretical Physics, Chinese Academy of Sciences, Beijing 100190, China}
\affiliation{School of Physical Sciences, University of Chinese Academy of Sciences, Beijing 100049, China}

\author{Su Yi}
\email{yisu@nbu.edu.cn}
\affiliation{Institute of Fundamental Physics and Quantum Technology and School of Physics Science and Technology, Ningbo University, Ningbo, 315211, China}

\author{Tao Shi}
\email{tshi@itp.ac.cn}
\affiliation{Institute of Theoretical Physics, Chinese Academy of Sciences, Beijing 100190, China}
\affiliation{School of Physical Sciences, University of Chinese Academy of Sciences, Beijing 100049, China}

\begin{abstract}
Dual-microwave shielding has emerged as a powerful tool for stabilizing ultracold polar molecules while tuning their intermolecular interactions. However, the two microwave fields are generally not perfectly orthogonal in experiments. Such misalignment introduces an in-plane component of the linearly polarized microwave, whose frequency differs from that of the elliptically polarized field. This component prevents complete cancellation of the dipole-dipole interaction and, more critically, renders the single-molecule dressed state intrinsically time-dependent, so that the conventional time-independent scattering framework is no longer available. Here we develop a Floquet theory that yields an analytic effective potential and enables accurate scattering calculations for polar molecules in non-orthogonal dual microwave fields. We find that, though misalignment weakens the shielding moderately, inelastic losses remain strongly suppressed under experimentally relevant conditions. Meanwhile, misalignment provides additional tunability of the interaction anisotropy and strength, which has been directly applied to recent experimental observations on the gas-to-droplet transition~[Z. Shi \textit{et al}, arXiv:2508.20518 (2025)] and Fermi-surface deformation in microwave-shielded molecular gases~[S. Biswas \textit{et al}, arXiv:2602.22447]. The framework is not restricted to dual-microwave shielding and can be generalized straightforwardly to arbitrary multi-frequency driving, providing a versatile tool for manipulating ultracold polar molecules under complex microwave configurations.

\end{abstract}

\maketitle
\section{Introduction}

Ultracold gases of polar molecules, characterized by their rich internal structure and long-range anisotropic dipole-dipole interaction (DDI), provide a versatile platform for quantum science~\cite{DeMille2002,Zoller2006a,
Zoller2006,Kozlov2007,Krem2008,Pfau2009,Carr2009,Berger2010,Hinds2011,Baranov2012,Bohn2017,Ye2017,Ni2019,Cornish2020,Zwierlein2021}. However, their application has long been limited by a nearly universal two-body loss mechanism that makes molecular gases unstable upon collisions~\cite{ye2010,Julienne2010,Julienne2011}. Microwave shielding addresses this difficulty by generating a repulsive shielding core between molecules and has dramatically suppressed collisional losses~\cite{Karman2018,Quemener2018,Doyle2021,Shi2023}. This progress has enabled recent breakthroughs, including degenerate Fermi gases
and Bose-Einstein condensates of microwave-shielded polar molecules (MSPMs) in NaK, NaCs, and NaRb systems~\cite{Luo2022a,Duda2023,Wang2023,Chen2023,Chen2024,Bigagli2023,Bigagli2024,Wang2025}. Among different realizations of MSPMs, the dual-microwave approach is particularly powerful: it was essential for the landmark realization of molecular condensates in NaCs~\cite{Bigagli2024,Zhang2026} and NaRb~\cite{Wang2025} by suppressing three-body losses, and it offers stronger shielding and greater interaction tunability than single-frequency schemes~\cite{Bigagli2024,Deng2025,Karman2025}.

In an ideal dual-microwave configuration, one microwave is circularly or elliptically polarized and propagates along the $z$ axis, with its field rotating in the $x$-$y$ plane, while the other is linearly polarized and propagates in the $x$-$y$ plane, with its field oscillating along the $z$ axis. The intermolecular interaction in this ideal setting is well described by an anisotropic effective potential consisting of a long-range DDI and a short-range repulsive core ($\propto1/r^6$)~\cite{Shi2023,Deng2025,zhang2025b}. This effective potential has become a standard tool for analyzing the few- and many-body physics of MSPMs~\cite{deng2024,jin2025,Langen2025,zhang2025a,Ciardi2025,zhang2025b,biswas2026}. However, in realistic experiments, the two microwave fields are inevitably misaligned, i.e., not exactly orthogonal~\cite{Bigagli2024,Wang2025,Zhang2026,biswas2026}. A finite tilt of the linearly polarized microwave field away from the $z$ axis introduces an additional linearly polarized component in the $x$-$y$ plane. This component has two consequences. First, it prevents exact cancellation of the DDI and leaves a residual long-range interaction that can modify both stability and evaporation. Second, the ground-to-rotational-excited-state coupling is driven simultaneously by the original circularly (or elliptically) polarized microwave and by the in-plane projection of the tilted linearly polarized field, generally at two different frequencies. Consequently, no rotating-frame transformation can render even the single-molecule Hamiltonian time independent. The dressed molecular state is intrinsically dynamical, which precludes the direct use of conventional stationary dressed states, effective potentials, and scattering theory.

Here, we develop a Floquet framework for polar molecules subjected to non-orthogonal dual microwave fields. The approach first defines the proper single-molecule dressed states in the multi-frequency field, indicating
that MSPMs exist in a dynamical dressed state, rather
than a stationary dressed state. By extending the construction to the two-body problem, we derive an analytic effective potential for two MSPMs, which serves as an accurate single-channel description of the interaction amenable to many-body physics. We further combine Floquet theory with multichannel scattering calculations to obtain scattering lengths and elastic and inelastic collision rates. The results reveal that misalignment can tune the effective interaction over a wide range, while the inelastic collision rate remains strongly suppressed for experimentally relevant parameters. The agreement between the full multichannel and single-channel calculations validates the effective potential and supports its use in many-body physics. More generally, the present Floquet construction is not limited to dual-microwave shielding and can be applied to molecular gases driven by arbitrary combinations of microwave frequencies.

\section{Non-ideal dual microwave fields}\label{model}

We consider polar molecules in the ${}^{1}\Sigma(v=0)$ state with electric dipole moment $d\hat{\mathbf{d}}$, where $\hat{\mathbf{d}}$ is the unit vector along the internuclear axis. The rotational Hamiltonian of a single molecule is $\hat{h}_{\mathrm{rot}}=B_{\mathrm{rot}}{\mathbf{J}}^{2}$, where $B_{\mathrm{rot}}$ is the rotational constant and ${\mathbf{J}}$ is the angular momentum operator. At ultracold temperatures, we can restrict the internal Hilbert space to the two lowest rotational manifolds, $J=0$ and $J=1$, split by $\hbar\omega_e=2B_{\mathrm{rot}}$. The relevant states are therefore $|J,M_J\rangle=|0,0\rangle$, $|1,0\rangle$, and $|1,\pm1\rangle$. Coupling between the two rotational manifolds is induced by microwave fields. We adopt the dual-microwave configuration used in recent experiments: one microwave propagates along the $z$ axis and is elliptically polarized with ellipticity $\xi$, so that its electric field rotates in the $x$-$y$ plane; the second microwave is linearly polarized, propagates approximately in the $x$-$y$ plane, and has a polarization axis tilted away from the $z$ axis by the polar and azimuthal angles $\vartheta_\pi$ and $\varphi_\pi$. The molecules are then illuminated by position-independent microwave fields,
\begin{widetext}
\begin{align}\label{EMW}
\mathbf E_\mathrm{MW}=&E_\sigma^+(\cos(\omega_\sigma t)\mathbf e_x+\sin(\omega_\sigma t)\mathbf e_y)
+E_\sigma^-(-\cos(\omega_\sigma t)\mathbf e_x+\sin(\omega_\sigma t)\mathbf e_y)-E_\pi^{(0)}\cos(\omega_\pi t)\mathbf e_z\nonumber\\
&+E_\pi^+(-\cos(\omega_\pi t+\varphi_\pi)\mathbf e_x-\sin(\omega_\pi t+\varphi_\pi)\mathbf e_y)
+E_\pi^-(-\cos(\omega_\pi t-\varphi_\pi)\mathbf e_x+\sin(\omega_\pi t-\varphi_\pi)\mathbf e_y),
\end{align}
\end{widetext}
Here, $\omega_\sigma$ and $\omega_\pi$ are the frequencies of the elliptically polarized and linearly polarized microwave fields. Equation~\eqref{EMW} follows from decomposing the fields into several components. The elliptically polarized microwave is a superposition of $\sigma^\pm$ components at frequency $\omega_\sigma$. The linearly polarized field contains one component along the $z$ axis and an in-plane component, which can itself be written as a superposition of two $\sigma^\pm$ components at frequency $\omega_\pi$. The corresponding amplitudes are
\begin{align*}
&E_\sigma^+ = E_\sigma \cos\xi, \quad E_\sigma^- = E_\sigma \sin\xi,\\
&E_\pi^+ = E_\pi^- = E_\pi \sin\vartheta_\pi, \quad E_\pi^{(0)} = E_\pi \cos\vartheta_\pi,
\end{align*}
where $E_\sigma$ and $E_\pi$ denote the amplitudes of the elliptically polarized and linearly polarized microwaves, respectively.

The coupling between the microwaves and the molecular rotational states gives rise to the Hamiltonian,
\begin{widetext}
\begin{align}
\hat h_\mathrm{MW}=&\frac{\hbar\Omega_\sigma^+}{2}e^{i\omega_\sigma t}|0,0\rangle\langle 1,1|
+\frac{\hbar\Omega_\sigma^-}{2}e^{i\omega_\sigma t}|0,0\rangle\langle 1,-1|\nonumber\\
&-\frac{\hbar\Omega_\pi^+}{2}e^{i(\omega_\pi t+\varphi_\pi)}|0,0\rangle\langle 1,1|
+\frac{\hbar\Omega_\pi^-}{2}e^{i(\omega_\pi t-\varphi_\pi)}|0,0\rangle\langle 1,-1|
+\frac{\hbar\Omega_\pi^{(0)}}{2}e^{i\omega_\pi t}|0,0\rangle\langle 1,0|
+\mathrm{h.c.},
\end{align}
\end{widetext}
where the Rabi frequencies are given by
\begin{align*}
&\Omega_\sigma^+ = \Omega_\sigma \cos\xi, \quad \Omega_\sigma^- = \Omega_\sigma \sin\xi,\\
&\Omega_\pi^\pm = \Omega_\pi \sin\vartheta_\pi, \quad \Omega_\pi^{(0)} = \Omega_\pi \cos\vartheta_\pi.
\end{align*}
Here, $\Omega_\sigma$ and $\Omega_\pi$ are the Rabi frequencies associated with the elliptically polarized and linearly polarized microwave fields.

We transform into a rotating frame using the unitary operator $\hat U(t)$,
\begin{align}
\hat{U}(t) =
\begin{pmatrix}
1 & 0 & 0 & 0 \\
0 & e^{-i\omega_\sigma t} & 0 & 0 \\
0 & 0 & e^{-i\omega_\pi t} & 0 \\
0 & 0 & 0 & e^{-i\omega_{\sigma} t}
\end{pmatrix}.
\end{align}
The choice of rotating frame is not unique, but different choices give equivalent physical results. In this frame, the internal-state Hamiltonian is
\begin{align}\label{single-molecule}
\hat h_\mathrm{in}=\sum_{n=-1}^1 \hat h_\mathrm{in}^{(n)}e^{in\omega t}
\end{align}
where \(\omega = \omega_\sigma - \omega_\pi\) is the frequency difference, and the matrices of $\hat h_\mathrm{in}^{(n)}$ are explicitly given by
\begin{align}
h_\mathrm{in}^{(0)}=&\left(\begin{array}{cccc}
               0 & \frac{\Omega_\sigma^+}{2} & \frac{\Omega_\pi^{(0)}}{2} & \frac{\Omega_\sigma^-}{2}\\
               \frac{\Omega_\sigma^+}{2} & \Delta_\sigma & 0 &0\\
               \frac{\Omega_\pi^{(0)}}{2} & 0 & \Delta_\pi & 0\\
               \frac{\Omega_\sigma^-}{2} & 0& 0& \Delta_\sigma
             \end{array}
\right),\\
h_\mathrm{in}^{(1)}=&
\left(\begin{array}{cccc}
               0 & 0 & 0 & 0\\
               -\frac{\Omega_\pi^+}{2}e^{-i\varphi_\pi} & 0 & 0 &0\\
               0 & 0 & 0 & 0\\
               \frac{\Omega_\pi^- e^{i\varphi_\pi}}{2} & 0& 0& 0
             \end{array}
\right),\\
h_\mathrm{in}^{(-1)}=&h_\mathrm{in}^{(1)\dag}
\end{align}
In the ideal case, $\vartheta_\pi=0$, so only $\hat h_\mathrm{in}^{(0)}$ remains and the Hamiltonian~\eqref{single-molecule} is time independent. Then, one can diagonalize $\hat h_\mathrm{in}$ to obtain four dressed states, $|\pm\rangle$, $|-1\rangle$, and $|0\rangle$, with eigenenergies $E_\pm$, $E_{-1}$, and $E_0$~\cite{Deng2025}. In practice, perfect alignment of the two microwaves is difficult and misalignment makes the Hamiltonian explicitly time-dependent. Stationary dressed states are then no longer available, which poses fundamental challenges for defining molecular interactions, scattering channels, and many-body Hamiltonian. In the following, we will demonstrate that this difficulty can be resolved within the framework of Floquet theory, which provides well-defined single-molecule states and a consistent route to the effective potential and scattering theory.

\section{Single molecule in microwave fields}

To define single-molecule states for the time-periodic Hamiltonian, we use Floquet theory. We start from the time-dependent Schr\"odinger equation for a single molecule,
\begin{align}\label{single-time-se}
i\hbar\frac{\partial|\phi(t)\rangle}{\partial t}=\hat h_\mathrm{in}|\phi(t)\rangle,
\end{align}
The solution can be written in the Floquet-Fourier form
\begin{align}\label{single-time-wave}
|\phi(t)\rangle=e^{-i\epsilon t/\hbar}\sum_{n=-\infty}^\infty e^{-in\omega t} |\phi_n\rangle
\end{align}
where $\epsilon$ is the quasienergy and $|\phi_n\rangle$ is the time-independent component in the $n$th Floquet harmonic. Substitution into Eq.~\eqref{single-time-se} gives the time-independent eigenvalue equation
\begin{align}\label{single-timeless-SE}
\sum_{s=-1}^1 \hat h_\mathrm{in}^{(s)}|\phi_{n+s}\rangle-n\omega|\phi_n\rangle=\epsilon|\phi_n\rangle,
\end{align}
We introduce the Floquet vector $|\boldsymbol{\Phi}\rangle=(\cdots,|\phi_1\rangle,|\phi_0\rangle,|\phi_{-1}\rangle,\cdots)^T$, which collects all harmonic components. Equation~\eqref{single-timeless-SE} then takes the compact form
\begin{align}\label{single-timeless-SE-matrix}
\mathbf h|\boldsymbol\Phi\rangle=\epsilon|\boldsymbol\Phi\rangle,
\end{align}
where the Floquet Hamiltonian $\mathbf h$ is a block-tridiagonal matrix with $4\times4$ blocks. Its eigenvectors are dressed states in the extended Hilbert space formed by the internal molecular states and the Floquet harmonics. This gives a consistent definition of single-molecule dressed states for arbitrary microwave fields. The ideal dual-microwave configuration is recovered when the off-diagonal Floquet blocks $h_\mathrm{in}^{(\pm1)}$ vanish; in that limit a single Floquet block is sufficient and one obtains the familiar stationary dressed states $|\pm\rangle$, $|-1\rangle$, and $|0\rangle$. Thus, in non-ideal microwaves, MSPMs should be regarded as dynamical Floquet-dressed molecules rather than conventional stationary dressed states.

Because these dressed states are superpositions over different Floquet harmonics, their labeling requires a convention. We denote by $|\mathbf F_\nu\rangle_n$ the single-molecule Floquet state that adiabatically connects, as $\vartheta_\pi\to0$, to the vector $|\boldsymbol\nu\rangle_n$. The latter has the internal state $|\nu\rangle$ in the $n$th harmonic and zero components in all other harmonics, with $|\nu\rangle=|\pm\rangle$, $|-1\rangle$, or $|0\rangle$. In microwave shielding experiments, molecules are prepared in $|\mathbf F_+\rangle_n$ to the shielding condition.

\section{Two molecules in microwave fields}

The well-defined Floquet dressed states provide a basis for treating molecular interactions and scattering. For two molecules with dipole moments $d\hat{\mathbf d}_1$ and $d\hat{\mathbf d}_2$, the inter-molecular DDI is
\begin{align}\label{DDI}
V_\mathrm{dd}=&\frac{d^2}{4\pi\epsilon_0 r^3}\left[ \hat{\mathbf d}_1\cdot\hat{\mathbf d}_2-3(\hat{\mathbf d}_1\cdot\hat{\mathbf r})(\hat{\mathbf d}_2\cdot\hat{\mathbf r})\right]\nonumber\\
=&-\frac{\eta}{r^3}\sum_{m=-2}^2 Y_{2m}^*(\hat{\mathbf r})\Sigma_{2,m}
\end{align}
where $\eta=\sqrt{8\pi/15}d^2/\epsilon_0$, $\epsilon_0$ is the vacuum permittivity, $\mathbf r$ is the relative coordinate of the two molecules, $r=|\mathbf r|$, $Y_{2m}(\hat{\mathbf r})$ are spherical harmonics, and $\Sigma_{2,m}$ are components of a rank-2 spherical tensor. Explicitly, $\Sigma_{2,0}=(\hat d_1^+\hat d_2^- + \hat d_1^-\hat d_2^+ + 2\hat d_1^0\hat d_2^0)/\sqrt6$, $\Sigma_{2,\pm1}=(\hat d_1^\pm\hat d_2^0 + \hat d_1^0\hat d_2^\pm)/\sqrt2$, and $\Sigma_{2,\pm2}=\hat d_1^\pm\hat d_2^\pm$, where $\hat d_j^\pm=Y_{1,\pm1}(\hat {\mathbf d}_j)$ and $\hat d_j^0=Y_{1,0}(\hat {\mathbf d}_j)$. In the rotating frame, the dipole components are
\begin{align}
\hat d^0=&\frac{1}{\sqrt{4\pi}}\left( |0,0\rangle\langle 1,0|e^{-i\omega_\pi t} +\mathrm{h.c.}\right),\\
\hat d^+=&\frac{1}{\sqrt{4\pi}}\left(- |0,0\rangle\langle 1,-1|e^{-i\omega_\sigma t}+|1,1\rangle\langle 0,0|e^{i\omega_\sigma t}\right),\\
\hat d^-=&-(\hat d^+)^\dag.
\end{align}
Substitution into $\Sigma_{2,m}$ gives
\begin{widetext}
\begin{align}
\Sigma_{2,0}=&\frac{1}{4\pi\sqrt 6}\left(2|1,0\rangle\langle 0,0|\otimes |0,0\rangle\langle 1,0|- |1,1\rangle\langle 0,0|\otimes|0,0\rangle\langle 1,1| - |0,0\rangle\langle 1,-1|\otimes|1,-1\rangle\langle 0,0|+\mathrm{h.c.}\right)\\
\Sigma_{2,1}
=&\frac{1}{4\pi\sqrt 2}\left[(|1,1\rangle\langle 0,0|\otimes|0,0\rangle\langle 1,0|+|0,0\rangle\langle 1,0|\otimes|1,1\rangle\langle 0,0|)e^{i\omega t}\right.\nonumber\\
&\left.-(|0,0\rangle\langle 1,-1|\otimes|1,0\rangle\langle 0,0|+|1,0\rangle\langle 0,0|\otimes|0,0\rangle\langle 1,-1|)e^{-i\omega t}\right]\\
\Sigma_{2,2}=&-\frac{1}{4\pi}\left(|1,1\rangle\langle 0,0|\otimes|0,0\rangle\langle 1,-1|+|0,0\rangle\langle 1,-1|\otimes|1,1\rangle\langle 0,0|\right)
\end{align}
\end{widetext}
Here we have made the rotating-wave approximation: terms oscillating at the microwave frequencies $\omega_\sigma$ or $\omega_\pi$ (typically GHz) are neglected, whereas terms oscillating at the much smaller difference frequency $\omega$ (typically MHz) are retained.

We now extend the single-molecule Floquet formalism to the two-body problem. The total Hamiltonian for the relative motion is
\begin{align}\label{two-SE}
\hat H_2(t)=-\frac{\hbar^2\nabla^2}{M}+\sum_{j=1,2}\hat h_\mathrm{in}(j)+V_\mathrm{dd}(\mathbf r,t),
\end{align}
where $M$ is the molecular mass and $\hat h_\mathrm{in}(j)$ is the time-dependent internal Hamiltonian of molecule $j$. Because $\hat H_2(t)$ possesses a parity symmetry, the symmetric and antisymmetric two-particle internal states are decoupled. We restrict the calculation to the ten-dimensional symmetric internal subspace $\mathcal S_{10}$ containing the microwave-shielded channel. Following the single-molecule construction, we write the solution of the time-dependent Schr\"odinger equation as
\begin{align}\label{two-time-wave}
|\psi(t)\rangle=e^{-i\mathcal{E} t/\hbar}\sum_{n=-\infty}^\infty e^{-in\omega t} |\psi_n\rangle
\end{align}
which leads to the eigenvalue equation
\begin{align}\label{two-time-independt-SE}
\sum_{s=-1}^1\mathcal H_s|\psi_{n+s}\rangle-n\omega|\psi_n\rangle=\mathcal E|\psi_n\rangle
\end{align}
where \(\mathcal{E}\) is the two-body quasi-energy and the Floquet Hamiltonian is
\begin{align*}
\mathcal H_s=\left\{
\begin{array}{ll}
 -\hbar^2\nabla^2/M+\sum_{j=1,2}\hat h_\mathrm{in}^{(0)}(j)+\mathcal V_0(r),&\text{for $s=0$} \\
\sum_{j=1,2}\hat h_\mathrm{in}^{(s)}(j)+\mathcal V_s(r),&\text{for $s=\pm1$}\\
0,&\text{otherwise}
\end{array}
\right.
\end{align*}
with the components satisfying $\mathcal V_{-1}=\mathcal V_{1}^\dag$. The explicit expressions of $\mathcal V$ are
\begin{align}\label{V_s}
\mathcal V_0(\mathbf r)=&-\frac{\eta}{r^3}\left[Y_{20}(\hat {\mathbf r})\Sigma_{20} +Y_{22}^*(\hat {\mathbf r})\Sigma_{22} +Y_{22}(\hat {\mathbf r})\Sigma_{22}^\dag \right],\\
\mathcal V_1=&-\frac{\eta}{r^3}\left[ Y_{21}^*(\hat{\mathbf r})\Sigma_{21}^{(1)} +Y_{21}(\hat{\mathbf r})\Sigma_{21}^{(-1)\dag} \right],
\end{align}
Here, $\Sigma_{21}^{(\pm1)}$ denotes the part of $\Sigma_{21}$ carrying the phase factor $e^{\pm i\omega t}$. Equation~\eqref{two-time-independt-SE} can again be written compactly as
\begin{align}
\mathbf H|\boldsymbol\Psi\rangle=\mathcal E|\boldsymbol\Psi\rangle,
\end{align}
where $|\boldsymbol\Psi\rangle=(\cdots,|\psi_1\rangle,|\psi_0\rangle,|\psi_{-1}\rangle,\cdots)^T$. The Floquet Hamiltonian $\mathbf H$ has the same block structure as the single-body Hamiltonian $\mathbf h$, but now includes the two-body internal space and the DDI. With the established Floquet eigenvalue equation~\eqref{two-time-independt-SE}, we can perform multichannel scattering calculations with the log-derivative algorithm~\cite{log-Johnson}.

To characterize the effective interaction between two molecules, we employ the Born-Oppenheimer approximation. For fixed intermolecular separation $\mathbf r$, we diagonalize the potential matrix $\mathbf V(\mathbf r)$, defined as the full Floquet Hamiltonian $\mathbf H$ without the kinetic-energy term, to obtain the adiabatic potentials
\begin{align}\label{adiabatic-potential}
\mathbf V(\mathbf r)|V^\mathrm{ad}(\mathbf r)\rangle=V^\mathrm{ad}(\mathbf r)|V^\mathrm{ad}(\mathbf r)\rangle,
\end{align}
where $V^\mathrm{ad}(\mathbf r)$ is the adiabatic potential and $|V^\mathrm{ad}(\mathbf r)\rangle$ is the corresponding adiabatic eigenstate. At large separation this eigenstate connects to an asymptotic two-body Floquet channel $|\mathbf F_\nu,\mathbf F_{\nu'}\rangle_n$. This state, in turn, adiabatically connects to $|\boldsymbol\nu,\boldsymbol\nu'\rangle_n$ in the limit $\vartheta_\pi\to0$. We focus on the experimentally relevant shielding channel $|\mathbf F_+,\mathbf F_+\rangle_n$.

Although the adiabatic potential $V^\mathrm{ad}(\mathbf r)$ gives a direct picture of the interaction, it is useful to derive an accurate effective potential $V_\mathrm{eff}(\mathbf r)$. This potential reduces the multichannel scattering problem to an efficient single-channel description while preserving the essential physics, and it also provides the input needed for many-body calculations. For $\vartheta_\pi\ne0$, the microwave configuration breaks both cylindrical symmetry about the $z$ axis and mirror symmetry with respect to the $x$-$y$ plane. Therefore, $V_\mathrm{eff}(\mathbf r)$ contains all allowed spherical harmonic components,
\begin{widetext}
\begin{align}\label{effective}
V_\mathrm{eff}(\mathbf r)=&\frac{1}{r^3}\left[C_{3,0}(3\cos^2\theta-1)+\sqrt\frac{2\pi}{15}C_{3,1}Y_{2,-1}(\theta,\phi)+\sqrt\frac{2\pi}{15}C_{3,-1}Y_{2,1}(\theta,\phi)\right.\nonumber\\
&\left.+\sqrt\frac{8\pi}{15}C_{3,2}Y_{2,-2}(\theta,\phi)+\sqrt\frac{8\pi}{15}C_{3,-2}Y_{2,2}(\theta,\phi)\right]
+\sum_{l=0,2,4}\sum_{|m|\le l}\frac{q_{l,-m}Y_{lm}(\theta,\phi)}{r^6},
\end{align}
\end{widetext}
The long-range coefficients $C_{3,m}$  take the analytic form of the time-averaged tensor components,
\begin{align}\label{C3}
C_{3,0}=&-8\sqrt\frac{2}{15}\pi^{3/2}d^2\sqrt\frac{5}{16\pi}\overline\Sigma_{2,0},\\
C_{3,1}=&8\sqrt\frac{2}{15}\pi^{3/2}d^2\sqrt \frac{15}{2\pi}\overline\Sigma_{2,1},\\
C_{3,2}=&-8\sqrt\frac{2}{15}\pi^{3/2}d^2\sqrt \frac{15}{8\pi}\overline\Sigma_{2,2},
\end{align}
where
\begin{align}
\overline\Sigma_{2,0}=&\sum_{n}\langle \psi_n|\Sigma_{2,0}|\psi_n\rangle,\\
\overline\Sigma_{2,1}
=&\sum_{n}\left(\langle \psi_{n-1}|\Sigma_{2,1}^{(1)}|\psi_n\rangle +\langle \psi_n|\Sigma_{2,1}^{(-1)}|\psi_{n-1}\rangle\right),\\
\overline\Sigma_{2,2}
=&\sum_{n}\langle \psi_n|\Sigma_{2,2}|\psi_n\rangle,
\end{align}
The short-range coefficients $q_{l,m}$ can be obtained from fitting the adiabatic potential numerically. The coefficients satisfy the condition, $C_{3,m}=(-1)^m C_{3,-m}^*$ and $q_{l,m}=(-1)^m q_{l,-m}^*$, ensuring that $V_\mathrm{eff}$ is real.

For specific microwave configurations, the effective potential simplifies. In the case $\xi=\varphi_\pi=0$, all coefficients are real and Eq.~\eqref{effective} reduces to
\begin{align}\label{effective0}
V_\mathrm{eff}(\mathbf r)=&\frac{1}{r^3}\left[C_{3,0}(3\cos^2\theta-1)+C_{3,1}\cos\phi\cos\theta\sin\theta+\right.\nonumber\\
&\left.C_{3,2}\cos(2\phi)\sin^2\theta\right]+\sum_{l=0,2,4}\sum_{|m|\le l}\frac{q_{l,-m}Y_{lm}}{r^6},
\end{align}
If $\xi=0$ but $\varphi_\pi\ne0$, the effective potential follows from Eq.~\eqref{effective0} by the rotation $V_\mathrm{eff}(\theta,\phi)\to V_\mathrm{eff}(\theta,\phi-\varphi_\pi)$. Equivalently, the coefficients transform as $C_{3,m}\to e^{im\varphi_\pi}C_{3,m}$ and $q_{l,m}\to e^{im\varphi_\pi}q_{l,m}$ in Eq.~\eqref{effective}. Thus, for $\xi=0$, $\varphi_\pi$ changes only the phases of the coefficients and not their magnitudes. In trapped ultracold gases, the interaction anisotropy, especially the direction of strongest attraction, can strongly affect the cloud shape~\cite{Wang2025,biswas2026}. For $\xi=0$, this direction is $(\theta_m,\varphi_\pi)$, with
\begin{align}
\theta_m = \frac{1}{2} \arctan\left( \frac{e^{-i\varphi_\pi}C_{3,1}}{3C_{3,0} - e^{-2i\varphi_\pi} C_{3,2}} \right).
\end{align}

We now specify the control parameters and units used in the numerical calculations. The molecular species enters through the permanent dipole moment $d$ and the mass $M$. The elliptically polarized microwave is characterized by the Rabi frequency $\Omega_\sigma$, detuning $\Delta_\sigma$, and ellipticity $\xi$, while the linearly polarized microwave is characterized by $\Omega_\pi$, $\Delta_\pi$, and the tilt angles $\vartheta_\pi$ and $\varphi_\pi$. We restrict $\xi\in[0,45^\circ]$ and $\vartheta_\pi,\varphi_\pi\in[0,90^\circ]$, since values outside these ranges can be mapped back by symmetry operations. Although the method applies to arbitrary microwave parameters, the numerical examples below focus on regimes close to cancellation of the long-range DDI. Since the repulsive shielding core prevents access to short intermolecular distances, we neglect the short-range van der Waals interaction. All results are expressed in dipolar units, with length $l_d=Md^2/(4\pi\epsilon_0\hbar^2)$, energy $E_d=\hbar^2/(Ml_d^2)$, and frequency $\omega_d=E_d/\hbar$. In these units the equations are species independent.

\subsection{Effective potential of two MSPMs}

The relative alignment of the two microwave fields controls both the strength and the anisotropy of the effective potential. In this section, we fix the Rabi frequencies and detunings in a weak-DDI regime and use the angles $\xi$, $\vartheta_\pi$, and $\varphi_\pi$ as tunable parameters. For $\xi=\vartheta_\pi=0$, the DDI is canceled in these Rabi frequencies and detunings. We characterize the DDI by the dipolar length~\cite{Chomaz_2022}
\begin{align}\label{dipole-sattering-length}
a_{3,m}=\frac{M}{3\hbar^2}C_{3,m}.
\end{align}
The key physics is the competition between the two microwave polarizations. The tilted linearly polarized field tends to enhance attraction along its polarization direction $(\vartheta_\pi,\varphi_\pi)$, whereas the elliptically polarized field favors attraction along the $y$ axis for $\xi>0$. Their interplay produces a strongly tunable anisotropic potential.

\begin{figure}[tbp]
\includegraphics[width=1\linewidth]{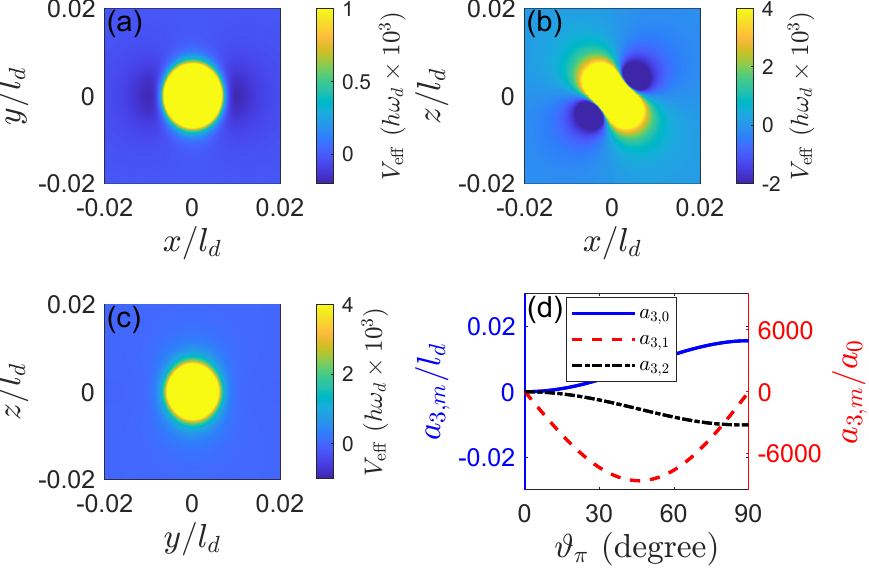}
\caption{Anisotropy of the effective potential between two microwave-shielded polar molecules. (a)--(c) Two-dimensional profiles in the $x$--$y$, $x$--$z$, and $y$--$z$ planes, respectively, for a polar angle $\vartheta_\pi = 10^\circ$. (d) Dipole lengths as functions of $\vartheta_\pi$. Parameters are fixed at $\varphi_\pi = 0$, $\xi = 0$, with Rabi frequencies and detunings being \(\Omega_\sigma = 2\pi \omega_d \times 6.0 \times 10^6\), \(\Omega_\pi = 2\pi \omega_d \times 5.3 \times 10^6\), \(\Delta_\sigma = 2\pi \omega_d \times 6.0 \times 10^6\), and \(\Delta_\pi = 2\pi \omega_d \times 8.0 \times 10^6\). An additional vertical axis in (d) is included assuming NaRb molecules and the corresponding $\omega_d$ is about $2$~Hz. Here, $a_0$ is Bohr radius.
}
\label{fig_effective0}
\end{figure}

We first consider $\xi=0$. In this case, a nonzero $\varphi_\pi$ only rotates the interaction, $V_\mathrm{eff}(r,\theta,\phi)\to V_\mathrm{eff}(r,\theta,\phi-\varphi_\pi)$, so it is sufficient to set $\varphi_\pi=0$. Figures~\ref{fig_effective0}(a)--(c) show two-dimensional profiles of the effective potential for $\vartheta_\pi=10^\circ$. The strongest attraction is neither in the $x$-$y$ plane nor along the $z$ axis; instead it is tilted within the $x$-$z$ plane. This tilt has direct experimental consequences, for example, the elongation of a molecular condensate along the most attractive direction~\cite{Wang2025}. Because the tilted $\pi$-polarized microwave has a projection along $x$, the coefficient satisfies $a_{3,2}\le0$, giving stronger attraction along $x$ than along $y$. Figure~\ref{fig_effective0}(d) shows that both $|a_{3,0}|$ and $|a_{3,2}|$ increase monotonically as $\vartheta_\pi$ is varied from $0$ to $90^\circ$. This follows from the weakening of the $z$ component and strengthening of the $x$ component of the linearly polarized field. In contrast, $a_{3,1}$ is non-monotonic: it vanishes at both $\vartheta_\pi=0$ and $90^\circ$ because mirror symmetry with respect to the $x$-$y$ plane is restored at both endpoints, and it reaches its maximum at an intermediate angle.

\begin{figure}[tbp]
\includegraphics[width=1\linewidth]{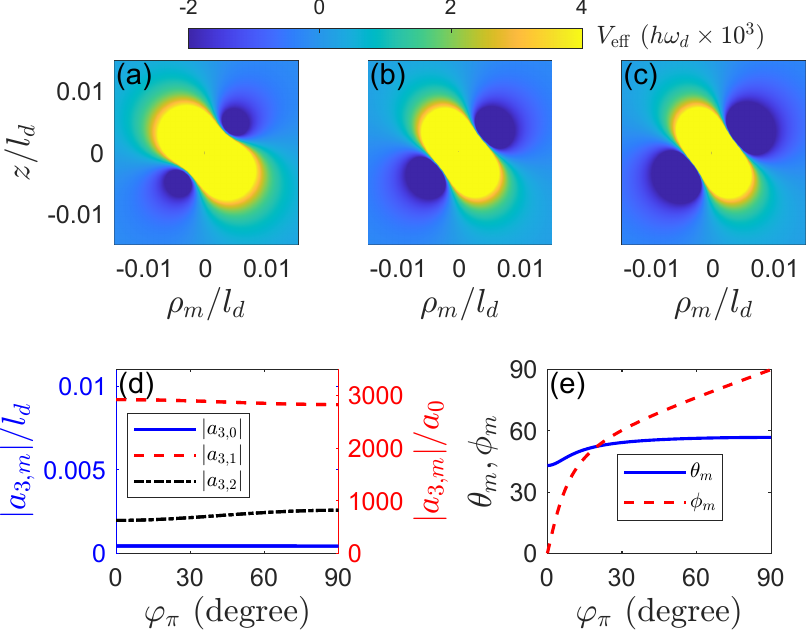}
\caption{
The effective potential with azimuthal angle $\varphi_\pi$, under weak ellipticity $\xi = 2^\circ$. (a)--(c) Potential in the $\rho_m$--$z$ plane at $\varphi_\pi = 0^\circ$ ($\phi_m = 0^\circ$), $30^\circ$ ($\phi_m = 60.2^\circ$), and $90^\circ$ ($\phi_m = 90^\circ$), respectively. (d) Variation of $|a_{3,m}|$ with $\varphi_\pi$. (e) Accompanying variation of the strongest attractive direction angles $\theta_m$ and $\phi_m$. The polar angle is fixed at $\vartheta_\pi = 10^\circ$; other parameters are as in Fig.~\ref{fig_effective0}. An additional vertical axis in (d) is included assuming NaRb molecules.
}
\label{fig_effective_xis2}
\end{figure}

\begin{figure}[tbp]
\includegraphics[width=1\linewidth]{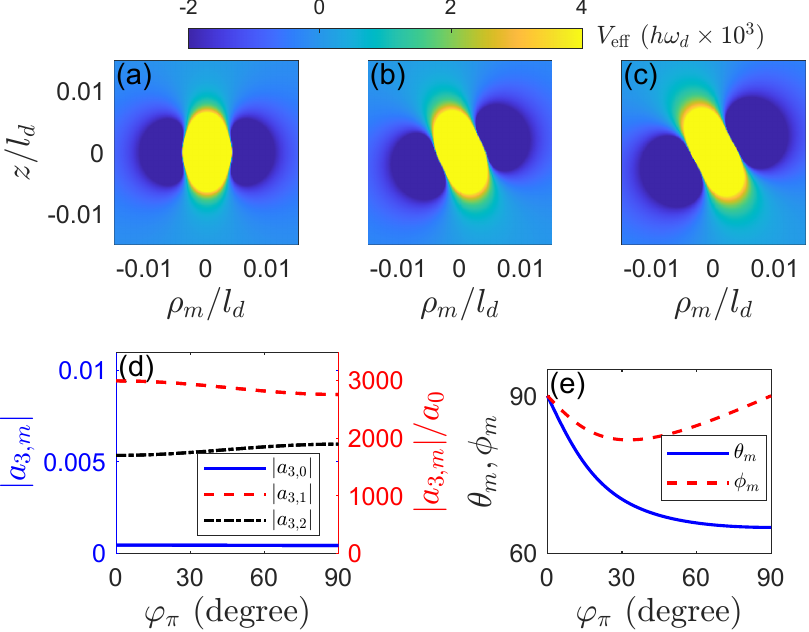}
\caption{
Effective potential under stronger ellipticity $\xi = 5^\circ$. (a)--(c) $\rho_m$--$z$ plane profiles for $\varphi_\pi = 0^\circ$ ($\phi_m = 90^\circ$), $30^\circ$ ($\phi_m = 81.6^\circ$), and $90^\circ$ ($\phi_m = 90^\circ$). (d) Dependence of $|a_{3,m}|$ on $\varphi_\pi$. (e) Variation of $\theta_m$ and $\phi_m$ with $\varphi_\pi$, showing non-monotonic behavior in $\phi_m$. Here, $\vartheta_\pi = 10^\circ$; remaining parameters are as in Fig.~\ref{fig_effective0}. An additional vertical axis in (d) is included assuming NaRb molecules.}
\label{fig_effective_xis5}
\end{figure}

For $\xi\ne0$, the competition between the two microwave polarizations induce complex behaviors of the effective potential. Figures~\ref{fig_effective_xis2}(a)--(c) show the effective potential in the $\rho_m$-$z$ plane for weak ellipticity, $\xi=2^\circ$, and a larger misalignment, $\vartheta_\pi=10^\circ$. Here $\hat{\boldsymbol\rho}_m=\cos\varphi_m\,\mathbf e_x+\sin\varphi_m\,\mathbf e_y$. The dependence on $\varphi_\pi$ is nontrivial: both the strength and the anisotropy of the effective potential change with $\varphi_\pi$. Nevertheless, the magnitudes $|a_{3,m}|$ vary only weakly with $\varphi_\pi$ [Fig.~\ref{fig_effective_xis2}(d)], and $|a_{3,0}|$ is almost unchanged because the $z$ projection of the linearly polarized field is fixed by $\vartheta_\pi$. The angles $\theta_m$ and $\varphi_m$ increase monotonically with $\varphi_\pi$ [Fig.~\ref{fig_effective_xis2}(e)], showing that, for these parameters, the most attractive direction mainly follows the tilted $\pi$-polarized field.

For stronger ellipticity the behavior changes, as shown in Fig.~\ref{fig_effective_xis5} for $\xi=5^\circ$ and $\vartheta_\pi=10^\circ$. At $\varphi_\pi=0$, the strongest attractive direction lies along the $y$ axis, and for $\varphi_\pi\ne0$ it remains close to this direction rather than following the tilted $\pi$-polarized field. The coefficients $|a_{3,m}|$ again depend only weakly on $\varphi_\pi$ [Fig.~\ref{fig_effective_xis5}(d)], but the angular dependence is qualitatively different: $\theta_m$ decreases monotonically, whereas $\varphi_m$ is non-monotonic and equals $90^\circ$ at both $\varphi_\pi=0$ and $90^\circ$. This behavior reflects the dominant role of the elliptically polarized microwave in setting the anisotropy.

\begin{figure}[tbp]
\includegraphics[width=1\linewidth]{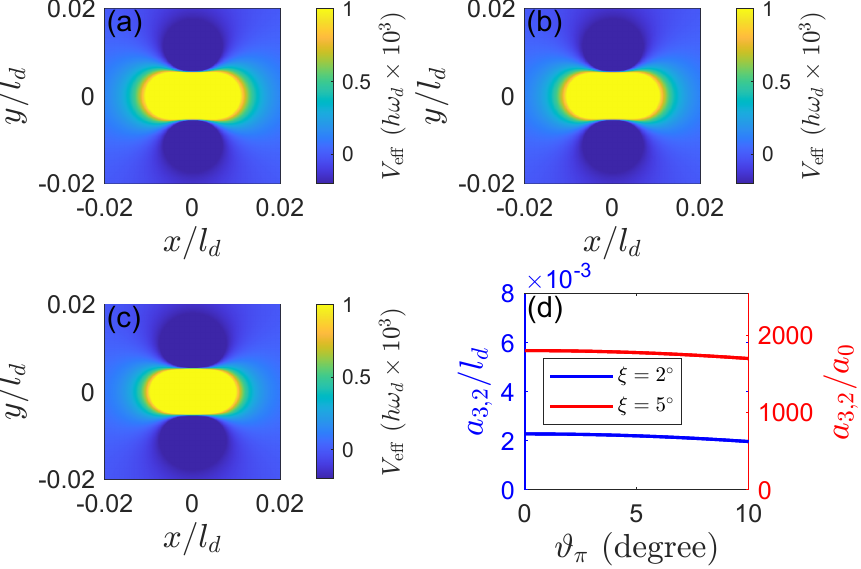}
\caption{The $x$--$y$ plane profiles of
the effective potential for (a) $\vartheta_\pi=0$, (b) $\vartheta_\pi=5^\circ$, and (c) $\vartheta_\pi=10^\circ$ with ellipticity $\xi=2^\circ$. (d) Dependence of $|a_{3,2}|$ on $\varphi_\pi$ for a small and large ellipticity. Here, $\varphi_\pi=0$; remaining parameters are as in Fig.~\ref{fig_effective0}. An additional vertical axis in (d) is included assuming NaRb molecules.}
\label{fig_effective_xy}
\end{figure}

Although misalignment can noticeably affect the potential in the $\rho_m$-$z$ plane near the DDI cancellation point, its contribution to the $x$-$y$ anisotropy is insignificant for typical experimental parameters, where $\vartheta_\pi$ does not exceed $\xi$ by orders of magnitude. In Figs.~\ref{fig_effective_xy}(a)--(c), the $x$-$y$ profile changes only weakly as $\vartheta_\pi$ increases from $0$ to $10^\circ$, even for the small ellipticity $\xi=2^\circ$. Consistently, $a_{3,2}$ depends only weakly on $\vartheta_\pi$ [Fig.~\ref{fig_effective_xy}(d)]. This is because the $x$-$y$ anisotropy induced by the tilted linearly polarized field is controlled by its in-plane projection, which is typically small compared with the elliptically polarized microwave contribution. Thus $a_{3,2}$, or equivalently $C_{3,2}$, is mainly determined by the ellipticity, explaining why a model without misalignment can still describe the observed radial Fermi-surface deformation~\cite{biswas2026}.

\subsection{Scattering of two MSPMs}

\begin{figure}[tbp]
\includegraphics[width=1\linewidth]{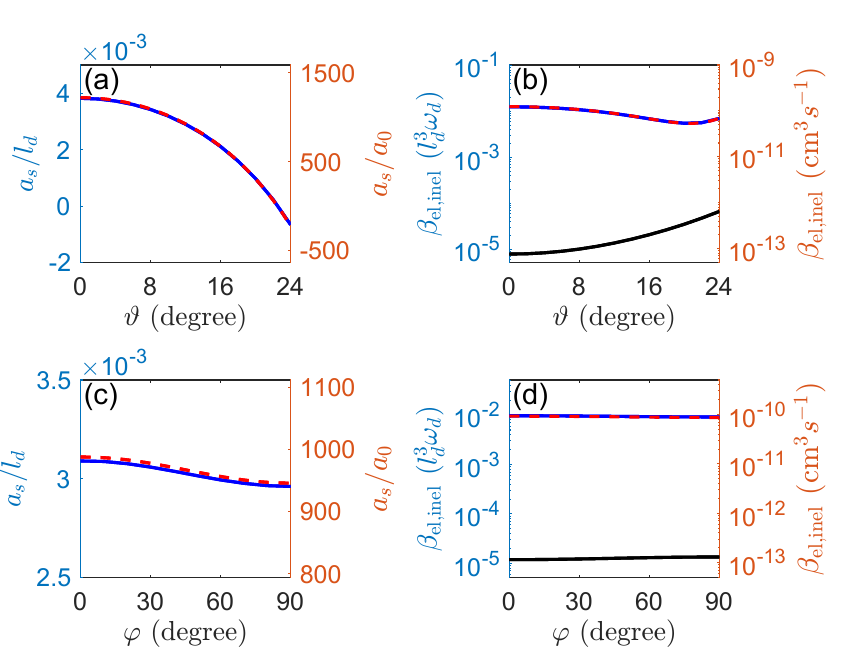}
\caption{Bosonic molecular scattering. (a) Scattering length and (b) elastic/inelastic rate versus $\vartheta_\pi$, when $\xi=\varphi_\pi=0$. (c) Scattering length and (d) elastic/inelastic rate versus $\varphi_\pi$, when $\xi=2^\circ$ and $\vartheta_\pi=10^\circ$. Solid lines denote multichannel calculations, while dashed lines denote single-channel calculations. The incident energy is $E_k=300E_d$, and other parameters match those in Fig.~\ref{fig_effective0}. Additional axes are provided for NaRb molecules.}
\label{fig_scattering_boson}
\end{figure}

Low-energy scattering is directly relevant to experiments and provides further validation of the effective potential. We first consider bosonic molecules in Fig.~\ref{fig_scattering_boson}, where the scattering length and elastic rates are compared between the full multichannel and single-channel calculations. The microwave frequencies and detunings are the same as in Fig.~\ref{fig_effective0}, so that $C_{3,m}$ vanishes when both ellipticity and tilt are absent. The close agreement over a broad parameter range confirms the validity of the effective potential. At $\vartheta_\pi=0$, only the repulsive core remains, and the $s$-wave scattering length is positive with a magnitude set by the radius of the shielding potential, typically of order $10^3a_0$. As $\vartheta_\pi$ increases, $a_s$ decreases and can become negative, indicating that misalignment makes the interaction more attractive. This tunability is large enough to drive the experimentally observed transition from a gas-phase condensate to a self-bound quantum droplet~\cite{Wang2025}. The inelastic rate also increases with $\vartheta_\pi$, but is still sufficiently small for microwave shielding to remain robust. Similar to the variation of $C_{3,m}$, the dependence of the scattering on $\varphi_\pi$ is weak in Figs.~\ref{fig_scattering_boson}(c) and (d), indicating that the net attraction undergoes only a slight change.

\begin{figure}[tbp]
\includegraphics[width=1\linewidth]{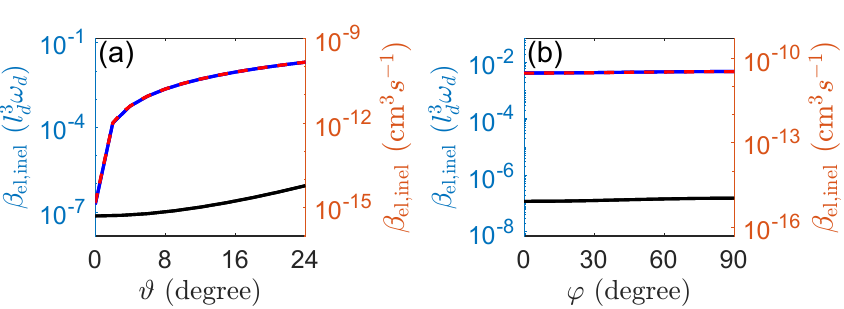}
\caption{Fermionic molecular scattering. (a) Elastic and inelastic rates as a function of $\vartheta_\pi$, with $\xi=\varphi_\pi=0$. (b) Elastic and inelastic rates versus $\vartheta_\pi$, for $\xi=2^\circ$ and $\varphi_\pi=10^\circ$. Solid lines denote multichannel calculations, while dashed lines denote single-channel calculations. The incident energy is set to $E_k=300E_d$; remaining parameters are consistent with Fig.~\ref{fig_effective0}. An additional vertical axis is provided for NaK molecules with the corresponding $\omega_d$ being about $21$~Hz.}
\label{fig_scattering_fermion}
\end{figure}

We next consider fermionic molecules in Fig.~\ref{fig_scattering_fermion}. The main trends parallel the bosonic case: multichannel and single-channel elastic rates agree well, the inelastic rate increases for nonzero $\vartheta_\pi$, and the dependence on $\varphi_\pi$ remains weak. The important difference is that identical fermions scatter dominantly in the $p$ wave, so the centrifugal barrier strongly suppresses short-range access. When the DDI is canceled, the remaining repulsive core is blocked by a centrifugal barrier, and the Fermi gases are nearly non-interacting [Fig.~\ref{fig_scattering_fermion}(a)]. Therefore, near the cancellation point, the strong suppression of the elastic rate renders evaporative cooling of fermionic molecules inefficient. Since Fermi gases remain stable under stronger DDI, it permits a straightforward improvement via tuning the Rabi frequencies, detunings, ellipticity, or misalignment to increase DDI and thereby enhance the elastic collision rate.

\section{Conclusion}

We have developed a Floquet framework for polar molecules subjected to non-orthogonal dual-microwave fields. The central complication is that microwave misalignment makes the single-molecule dressed state intrinsically time-dependent, invalidating the usual stationary dressed-state construction. By working in the extended Floquet space, we define the relevant dynamical dressed states, formulate the corresponding two-body scattering problem, and derive an analytic effective potential for MSPMs. The resulting DDI coefficients quantify how misalignment generates long-range interactions and changes the direction of strongest attraction. Full multichannel scattering calculations show that, although misalignment can weaken shielding and modify the elastic rate, the inelastic loss rate remains strongly suppressed in experimentally relevant regimes. The close agreement between the full multichannel and single-channel calculations validates the effective potential as a compact description of the two-body interaction and its extension to many-body physics. In particular, the residual attractive interaction produced by microwave misalignment accounts for the tunability used in the observed gas-to-droplet transition of NaRb molecules~\cite{Wang2025}, while the ellipticity-controlled in-plane anisotropy underlies the controlled Fermi-surface deformation reported in microwave-shielded molecular gases~\cite{biswas2026}. More broadly, the Floquet construction offers a systematic route to molecular gases driven by arbitrary multi-frequency microwave fields.

\section{Acknowledgment}
We thank W. Zhang, Z. Huang, and Z. Shi for fruitful discussions. This work was supported by the National Key Research and Development Program of China (Grant No. 2021YFA0718304) and by the National Natural Science Foundation of China (Grants No. 12504313, No. 12135018, No. 12525413, and No. 12574295).

\bibliography{ref_unideal_microwave}

\end{document}